\documentclass[journal]{IEEEtran}
\usepackage{graphicx}
\usepackage{tipa}
\usepackage{bbm}
\usepackage{mathrsfs}
\usepackage{amsfonts}
\usepackage{cite,url,subfigure,epsfig,graphicx}
\usepackage{amssymb,amsmath}
\usepackage{indentfirst}
\usepackage{algorithmic}
\usepackage{algorithm}
\usepackage{pdfpages}
\usepackage{pifont}
\usepackage{cases}
\usepackage{cite,url,subfigure,epsfig,graphicx}
\usepackage{times,verbatim,amsfonts,amsmath,color}

\IEEEoverridecommandlockouts
\makeatletter
\def\ps@headings{\def\@oddhead{\mbox{}\scriptsize\rightmark \hfil \thepage}\def\@evenhead{\scriptsize\thepage \hfil \leftmark\mbox{}}\def\@oddfoot{}\def\@evenfoot{}}
\makeatother 

\newtheorem{definition}{ \textbf{Definition}}
\newtheorem{lemma}{ \textbf{Lemma}}

\begin{document}

\title{Trade Privacy for Utility: A Learning-Based Privacy Pricing Game in Federated Learning}
\author{
\IEEEauthorblockN{Yuntao~Wang\IEEEauthorrefmark{2}, Zhou~Su\IEEEauthorrefmark{2}, Yanghe~Pan\IEEEauthorrefmark{2}, Abderrahim~Benslimane\IEEEauthorrefmark{3}, Yiliang~Liu\IEEEauthorrefmark{2}, Tom~H.~Luan\IEEEauthorrefmark{2}, and Ruidong~Li\IEEEauthorrefmark{4}}\\
\IEEEauthorblockA{
\IEEEauthorrefmark{2}School of Cyber Science and Engineering, Xi'an Jiaotong University, China\\
\IEEEauthorrefmark{3}Laboratory of Computer Sciences, Avignon University, France\\
\IEEEauthorrefmark{4}College of Science and Engineering, Kanazawa University, Japan\\
Corresponding author: zhousu@ieee.org
}}
\maketitle

\begin{abstract}
To prevent implicit privacy disclosure in sharing gradients among data owners (DOs) under federated learning (FL), differential privacy (DP) and its variants have become a common practice to offer formal privacy guarantees with low overheads. However, individual DOs generally tend to inject larger DP noises for stronger privacy provisions (which entails severe degradation of model utility), while the curator (i.e., aggregation server) aims to minimize the overall effect of added random noises for satisfactory model performance.
To address this conflicting goal, we propose a novel dynamic privacy pricing (DyPP) game which allows DOs to sell individual privacy (by lowering the scale of locally added DP noise) for differentiated economic compensations (offered by the curator), thereby enhancing FL model utility.
Considering multi-dimensional information asymmetry among players (e.g., DO's data distribution and privacy preference, and curator's maximum affordable payment) as well as their varying private information in distinct FL tasks, it is hard to directly attain the Nash equilibrium of the mixed-strategy DyPP game.
Alternatively, we devise a fast reinforcement learning algorithm with two layers to quickly learn the optimal mixed noise-saving strategy of DOs and the optimal mixed pricing strategy of the curator without prior knowledge of players' private information.
Experiments on real datasets validate the feasibility and effectiveness of the proposed scheme in terms of faster convergence speed and enhanced FL model utility with lower payment costs.
\end{abstract}


\IEEEpeerreviewmaketitle

\section{Introduction}
Driven by the unprecedented amount of data generated by smart devices, recent years have witnessed the exciting advances of artificial intelligence (AI), especially deep learning, for a wide range of smart applications such as smart surveillance and machine translation. In the traditional centralized AI paradigm, data across various data owners (DOs) such as mobile users should be concentrated for data mining and model training \cite{9579038}, which raises severe privacy breaches. 
Federated learning (FL) is a distributed AI paradigm which allows DOs to collaboratively train a shared AI model without disclosing their local private data \cite{9928220,9664267}. In a typical FL system, DOs periodically send the intermediate gradients (i.e., local model update) computed on local datasets to the curator (which synthesises a global model). Then, the curator distributes the updated global model back to DOs for next-round distributed on-device learning. This procedure is repeated until achieving a desirable accuracy of the global model. Under FL, DOs' private data are kept on local devices, thereby greatly mitigating user privacy concerns.

Nevertheless, evidences have demonstrated that FL can be susceptible to advanced inference attacks such as membership inference attacks \cite{NEURIPS2019,8835245,Hitaj2017CCS} (that infer whether a particular data sample is involved in a DO's private training dataset) and model reconstruct attacks \cite{Fredrikson2015CCS,8835269} (that recover DOs' private training data) in exchanging intermediate gradients. To enhance privacy protection, differential privacy (DP) methods \cite{9760102,9740410,9790807} have become a common practice in FL due to the low overheads and rigorous privacy guarantees, where DOs independently obfuscate local model updates by adding artificial DP noises.
However, in practical DP-based FL applications, self-interested DOs tend to inject larger random noises to enforce stronger privacy provisions, which eventually entails severe degradation of model utility. By contrast, the curator aims to minimize the overall effect of injected DP noises for satisfactory model performance. Additionally, DOs typically exhibit heterogeneous privacy expectations \cite{7093125}. For example, DOs can have distinct sensitivity levels towards potential privacy leakage under the uniform privacy protection level (PPL).
Therefore, it necessities a privacy-utility tradeoff in FL while satisfying DOs' customized privacy expectations.

As an attempt to address this issue, Sun \emph{et al}. \cite{9565851} recently introduced a personalized privacy pricing approach named Pain-FL, which offers differentiated PPL-payment contracts for DOs with customized privacy expectations in exchange of their reduced scale of locally added DP noises. 
In Pain-FL \cite{9565851}, each DO selects an optimal PPL-payment contract and perturbs its local model update with that PPL in exchange for the corresponding payment. Accordingly, only moderate amount of DP noises is added by DOs, thereby ensuring satisfactory model performance.

However, there are still significant challenges remaining to be resolved. 1) The design of optimal contracts in \cite{9565851} requires precise distribution information of DOs' privacy types, which can be non-trivial in practice. For example, the same DO may have distinct privacy requirements when undertaking different FL tasks involving different local private data; meanwhile, DO's privacy preferences may change over time. 2) Pain-FL \cite{9565851} only considers DOs' diverse privacy types in contract design while their multi-dimensional private information (e.g., local data size and distribution) is neglected, which inevitably deteriorates the contract efficiency. 
3) As the optimal contracts are centrally designed by the curator in \cite{9565851} and only support pure strategy, it lacks contract adaptability and feasibility under distributed and mixed-strategy contract design settings.

To this end, this paper proposes a novel reinforcement learning (RL)-based privacy pricing scheme to intelligently learn the optimal privacy trading strategies (i.e., injected DP noise scale strategy of DOs and pricing strategy of the curator) without the reliance on prior knowledge of players' private information.
Specifically, we first formulate the interactions between DOs and the curator as a dynamic privacy pricing (DyPP) game with mixed strategy. In DyPP game, the curator determines privacy pricing strategies for heterogeneous DOs with multi-dimensional private information (i.e., privacy cost, training data size, and data distribution); while each DO determines the amount of traded privacy by varying the variance of locally injected Gaussian noise. 
To derive the Nash equilibrium (NE) of the game under the dynamic and uncertain environment with multi-dimensional information asymmetry, we also devise a fast RL algorithm with two tiers, by leveraging Win or Learn Fast Policy Hill-Climbing (WoLF-PHC) methods, to quickly search the optimal mixed-strategy policies for both DOs and the curator.
Finally, extensive experiments demonstrate that the proposed scheme can fast converge to the NE and improve model utility with lower payments.

The rest of the paper is organized as follows. Related works are reviewed in Section~\ref{sec:RELATEDWORK}. The system model and DyPP game are formulated in Section~\ref{sec:SYSTEMMODEL}. Section~\ref{sec:FRAMEWORK} presents the two-layer RL-based game solution. Section~\ref{sec:SIMULATION} gives the performance evaluation. Section~\ref{sec:CONSLUSION} concludes this paper.

\section{Related Works}\label{sec:RELATEDWORK}
Recently, a number of efforts have been reported to seek a privacy-utility balance in FL from the perspective of economic incentives, and many of them are based on the contract theory.
Saputra \emph{et al}. \cite{9585537} formulated optimal payment contracts for vehicles with diverse quality of sensing information in FL-based vehicular crowdsensing scenarios to maximize the FL platform's profits under the payment budget.
By leveraging contract theory, Sun \emph{et al}. \cite{9565851} proposed a customized contract-based market model to incentivize workers with distinct privacy preferences to participate in FL, where the contract specifies worker's PPL and the corresponding payment in every learning round.
Ding \emph{et al}. \cite{9252911} investigated the optimal contract design for participants with 2D private information (i.e., training costs and communication delay) in wireless networks under three information asymmetry levels.
Lim \emph{et al}. \cite{9057543} presented a hierarchical incentive mechanism in FL to address the incentive mismatches between model owners and data owners, as well as among model owners. A contract theoretical approach is devised to recruit qualified data owners to collaboratively train the FL model, and a coalitional game approach is designed to allocate model profits according to the marginal contributions.

However, the working of the above works relies on the prior knowledge of DOs' private information and none of them consider the multi-dimensional private information including privacy types, training data sizes, and non-IID degrees in optimal contract design in distributed and mixed-strategy settings.

\section{System Model and Game Formulation}\label{sec:SYSTEMMODEL}
\begin{figure}[!t]\setlength{\abovecaptionskip}{-0.1cm}\vspace{-1.mm}
\centering
  \includegraphics[width=7.8cm,height=6.3cm]{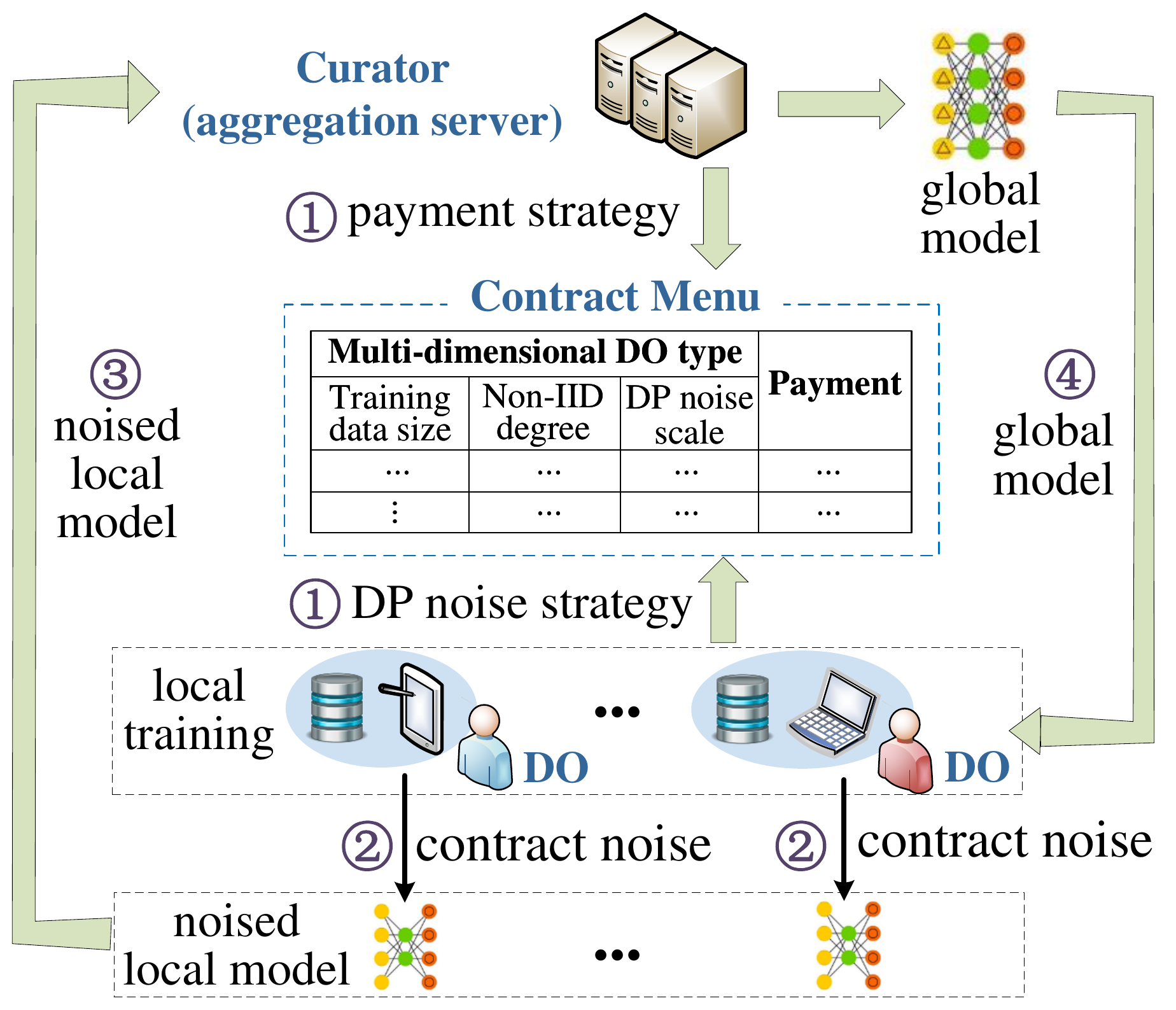}
  \caption{Illustration of the dynamic privacy pricing (DyPP) game in FL.}\label{fig:model}\vspace{-4mm}
\end{figure}
\subsection{Framework Overview}\label{subsec:networkmodel}
Fig.~\ref{fig:model} illustrates our proposed FL framework, consisting of the curator (denoted as $S$) and a set of $N$ individual DOs (denoted as $\mathcal{N} = \{1,\cdots,N\}$). In FL, DOs in $\mathcal{N}$ collaboratively train a shared AI model using local datasets, coordinated by the curator $S$ which serves as the aggregation server. Typically, DOs in $\mathcal{N}$ have diverse training data sizes, data distributions, and privacy leakage costs, which are their private information.
As shown in Fig.~\ref{fig:model}, the workflow of our proposed framework contains the following steps.
\begin{enumerate}
  \item DOs and the curator jointly design a series of personalized contracts, where each contract bundle specifies the relation among the payment $p_n$, privacy cost $c_{n}$, training data size $D_n$, and non-IID degree $\beta_n$ (step \ding{172}). Specifically, each DO $n \in \mathcal{N}$ independently determines his customized DP noise-saving strategy $\Delta\sigma_n$. In conventional uniform DP, the uniform level of added DP noise is usually set at a relatively large value (i.e., $\sigma_{\max}$) to satisfy the privacy needs of most DOs.
      Here, $\Delta\sigma_n = \sigma_{\max} - \sigma_{n}$ means the saved noise scale of DO $n$, where $\sigma_{n}$ denotes DO $n$'s scale of added Gaussian noise. Meanwhile, the curator $S$ determines its payment strategy $p_n$ to compensate for each DO $n$'s privacy loss.
  \item After both sides complete the contract design and signing process, at each communication round $i$ ($1\le i \le I$), each DO $n \in \mathcal{N}$ trains the global model $\widetilde{\Psi}^{i-1}$ using mini-batch SGD with batch size $X_n$. Then, DO $n$ perturbs his locally computed gradients (i.e., local model) $\Psi_n^i$ based on the noise scale $\sigma_{n}$ specified by the signed contract item (step \ding{173}).
  \item DO $n \in \mathcal{N}$ uploads the noised version of local model (i.e., $\widetilde{\Psi}_n^i$) to the curator $S$ (step \ding{174}).
  \item The curator $S$ synthesises a global model $\widetilde{\Psi}^i$ by aggregating all the noised local models and delivers $\widetilde{\Psi}^i$ to each DO for next-round training (step \ding{175}).
  \item After the global model achieves a desirable accuracy or the learning round $i$ attains its maximum value $I$, the learning process ends and the curator delivers the contractual payment $p_n$ to every DO $n$.
\end{enumerate}

\subsection{Mixed-Strategy Dynamic Privacy Pricing Game}\label{subsec:gamemodel}
We employ the well-known zero-concentrated DP (zCDP) \cite{924681}, as a relaxation of DP with tight composition bound, to facilitate privacy and convergence analysis in FL with customized privacy provisions, as shown in Definition 1. 
\begin{definition}[$\alpha$-zCDP]
For any $D,D'\in \mathcal{X}^{d}$ that differ on a single data sample and any $\varphi>1$, a randomized mechanism $\mathcal{M}: \mathcal{X}^{d} \rightarrow \mathcal{Y}$ satisfies $\alpha$-zCDP if
\begin{align}\label{eq:zCDP}
\mathbb{E}[\exp\left((\varphi-1)\Upsilon\right)] \leq \exp\left((\varphi-1)(\alpha \cdot \varphi)\right),
\end{align}
where $\mathbb{E}$ indicates the expectation operator. $\alpha$ is the privacy budget, and a smaller $\alpha$ enforces a larger PPL. $\Upsilon$ is the privacy loss random variable \cite{924681}, implying the likelihood between $D$ and $D'$ given $\mathcal{M}(D)$ or $\mathcal{M}(D')$. The larger $\Upsilon$ means the larger likelihood ratio (or the weaker privacy protection). 
\end{definition}

\begin{lemma}\label{lemma1}
The Gaussian mechanism $\mathcal{M}$ satisfies $(\Delta_f^2/2\sigma^2)$-zCDP by adding artificial noise following Gaussian distribution $\mathbb{N}(0,\sigma^2\mathrm{I}_d)$, where $\Delta_f$ is the query sensitivity, $\sigma^2$ is the noise variance, and $\mathrm{I}_d$ is a $d$-dimensional identity matrix. 
\end{lemma}

\begin{IEEEproof}
Please refer to Proposition 1.6 in \cite{924681}.
\end{IEEEproof}

Next, we formulate a dynamic privacy pricing (DyPP) game with mixed strategy in FL, where the curator and DOs can dynamically randomize their strategies to fool the opponent.

\begin{definition}[Mixed-Strategy DyPP Game]
The interactions between the curator and DOs can be formulated as a DyPP game with mixed strategy, i.e., $\mathcal{G}=\{S,\mathcal{N},\{\{p_n,\Delta\sigma_n\}_{n\in \mathcal{N}}\},\{\mathbf{x}_n,\mathbf{y}_n\}_{n\in \mathcal{N}},\{\mathcal{F}_S,\{\mathcal{F}_n\}_{n\in \mathcal{N}}\}\}$.
\begin{itemize}
  \item \emph{Player.} The curator $S$ and a group of DOs in $\mathcal{N}$ are players in the game $\mathcal{G}$.
  \item \emph{Mixed-Strategy.} The curator $S$ quantizes its pricing strategy into $K\!+\!1$ levels, i.e., $p_n\in \{\frac{k}{K\cdot p_{\max}}\}_{0\leq k \leq K}$ and decides the mixed pricing strategy, i.e.,
      \begin{numcases}{}
      \mathbf{x}_n = [x_{n,k}]_{0\leq k \leq K}\in \Pi, \hfill\\
      x_{n,k} = \Pr\big(p_n = \frac{k}{K\cdot p_{\max}}\big),
      \end{numcases}
      to compensate for DO's privacy loss while maximizing its overall payoff. $\Pi$ denotes the curator's policy set, and $p_{\max}$ is the maximum affordable payment.
      Each DO $n \in \mathcal{N}$ quantizes his DP noise-saving strategy into $J+1$ levels, i.e., $\Delta\sigma_n\in \{\frac{j}{J\cdot \sigma_{\max}}\}_{0\leq j \leq J}$, and determines the mixed noise-saving strategy for optimized payoff, i.e.,
      \begin{numcases}{}
      \mathbf{y}_n = [y_{n,j}]_{0\leq j \leq J}\in \Psi, \hfill\\
      y_{n,j} = \Pr\big(\Delta\sigma_n = \frac{j}{J\cdot \sigma_{\max}}\big),
      \end{numcases}
      where $\Psi$ is DO $n$'s policy set. We have $x_{n,k},y_{n,j}\geq 0$ and $\sum_{k=0}^{K}{x_{n,k}} = \sum_{j=0}^{J}{y_{n,j}} = 1$ by definition.
  \item \emph{Payoff.} Let $\mathcal{F}_S$ and $\mathcal{F}_n$ denote the expected payoffs of the curator $S$ and each DO $n$, respectively.
\end{itemize}
\end{definition}

\underline{Expected Payoff of DO.} The expected payoff of DO $n \in \mathcal{N}$ is denoted as the revenue minuses the privacy loss:
\begin{align}\label{eq:payoffDO}
\mathcal{F}_{n}&\left(\mathbf{x}_n, \mathbf{y}_n\right) = \nonumber\\
&\sum_{k=0}^{K} {\sum_{j=0}^{J} {x_{n,k}\,y_{n,j} \big[\lambda_r\cdot p_n - \nu \cdot c_n \left(\sigma_{\max} - \Delta\sigma_n\right) \big]}},
\end{align}
where $\lambda_r, \nu$ are positive adjustment factors. $c_n$ is DO $n$'s unit privacy leakage cost, which is secret to others. The last term in Eq. (\ref{eq:payoffDO}) denotes the privacy loss of DO $n$, which is related to the scale of added Gaussian noise.
According to \cite{9565851}, the query sensitivity of DO $n$'s local model $\Psi_n^i$ is $\Delta_f = \frac{2L}{X_n}$, where $L$ is the Lipschitz constant. Based on Lemma~\ref{lemma1}, the Gaussian mechanism meets $\alpha_n$-zCDP with $\alpha_n = \frac{2L^2}{X_n^2 \sigma_n^2}$.

\underline{Expected Payoff of Curator.} The expected payoff of the curator $S$ contains two parts: the overall quality of the aggregated global model and the accumulated payment to DOs, i.e.,
\begin{align}\label{eq:payoffCurator}
&\mathcal{F}_{S}\left(\mathbf{x}, \mathbf{y}\right) = \sum\nolimits_{n\in \mathcal{N}} \mathcal{F}_{S,n}\left(\mathbf{x}_n, \mathbf{y}_n\right)=\nonumber\\
&\sum_{n=1}^{N} {\sum_{k=0}^{K} {\sum_{j=0}^{J} {x_{n,k}\,y_{n,j} \big[\varpi \lambda_s \mathcal{A}(\Delta\sigma_n,\beta) \!-\! (1 \!-\! \varpi) \mu \cdot p_n \big]}}},
\end{align}
where $\varpi \in [0,1]$ is the weight parameter, indicating the curator's sensitivity to model quality. $\mu > 0$ is an adjustment factor. $\mathbf{x} = [\mathbf{x}_n]_{n\in\mathcal{N}}$ and $\mathbf{y} = [\mathbf{y}_n]_{n\in\mathcal{N}}$. For simplicity, the quality of the aggregated global model is evaluated via the sum of quality of DOs' local models \cite{7093125}. In Eq. (\ref{eq:payoffCurator}), $\mathcal{A}(.)$ is the quality function measured by the model loss ${\mathscr{L}}(.)$, i,e.,
\begin{align}\label{eq:modelQuality}
\mathcal{A}(\Delta\sigma_n,\beta) = - \zeta_1 \cdot {\mathscr{L}}(\Delta\sigma_n,\beta) + \zeta_2,
\end{align}
where $\zeta_1, \zeta_2>0$ are adjustment factors. $\zeta_2$ denotes the maximum model quality when ${\mathscr{L}}\rightarrow 0$.
From Eq. (\ref{eq:modelQuality}), the smaller the model loss, the higher the local model quality.

DOs generally have distinct privacy preferences (by adding distinct Gaussian noises on local models), training data sizes and data distributions, resulting in distinct quality of uploaded local models. Based on \cite{Hsu2019MeasuringTE}, the Dirichlet distribution characterizes DOs' heterogeneity in terms of data size and data distribution. DO's training examples in a typical $Y$-class classification task are drawn from a Dirichlet distribution $\mathrm{Dir}({\beta})$, where ${\beta}$ captures the non-IID degree. 
Especially, $\beta \rightarrow 0$ means DOs only randomly have one class of samples, while $\beta \rightarrow \infty$ is the IID case.
Based on experimental validations in our previous work \cite{corr/abs-2212-13992} (i.e., Figs.~4--7), the model loss function ${\mathscr{L}}(.)$ can be modeled as a 3D sigmoid curve with the non-IID degree ${\beta}$ and the saved noise scale $\Delta\sigma_{n}$, i.e.,
\begin{align}\label{eq:lossfitting}
{\mathscr{L}}(\Delta\sigma_n,\beta) = \frac{\gamma_1 \exp(-\gamma_2 \cdot \beta)}{\gamma_3 + \exp(-\gamma_4 \left(\sigma_{\max} - \Delta\sigma_n\right))} + \gamma_5,
\end{align}
where $\gamma_1,\cdots,\gamma_5>0$ are curve-fitting parameters. From Eq. (\ref{eq:lossfitting}), a higher non-IID degree $\gamma$ results in a diminishing marginal model loss, and a larger saved noise scale $\Delta\sigma_{n}$ entails a performance enhancement.

In the mixed-strategy DyPP game, we have two conflicting goals. Particularly, the curator tends to minimize the variance of totally added Gaussian noises for satisfactory model performance with a low payment, while DOs tend to add Gaussian noises with higher variances to pursue stronger privacy provisions.
The solution of the game is the Nash equilibrium (NE), in which no player can improve his payoff by unilaterally
deviating from it \cite{9696188}.
The NE of the mixed-strategy DyPP game is denoted as
\begin{align}\label{eq:NE}
\mathcal{F}_{S}\left(\mathbf{x}^*, \mathbf{y}^*\right) &\geq \mathcal{F}_{S}\left(\mathbf{x}, \mathbf{y}^*\right),\forall \mathbf{x} \in \Pi^N,\\
\mathcal{F}_{n}\left(\mathbf{x}_n^*, \mathbf{y}_n^*\right) &\geq \mathcal{F}_{n}\left(\mathbf{x}_n^*, \mathbf{y}_n\right), \forall n \in \mathcal{N}, \forall \mathbf{y}_n \in \Psi.
\end{align}

\section{Two-Layer RL-Based DyPP Game Solution}\label{sec:FRAMEWORK}
%

Due to the existence of multi-dimensional information asymmetry, the curator is usually unaware of the distribution of privacy cost (or privacy preference) among DOs, while DOs are usually unaware of the curator's payment model (e.g., maximum affordable payment and sensitivity to model quality). Besides, the preferences of the curator and DOs may vary under dynamic and uncertain environments. For instance, when undertaking different FL missions, a DO can have distinct privacy expectations and the curator can have distinct sensitivities to model quality.
Under such strong information asymmetry scenarios, the curator and DOs can separately employ the WoLF-PHC algorithm (a model-free RL method) to derive the optimal policy in the mixed-strategy DyPP game under dynamic and uncertain environments.

\subsection{Intelligent Noise-Saving Strategy Based on WoLF-PHC}\label{subsec:scheme1}
For each DO, his noise-saving strategy-making process in repeated interactions can be formulated as a finite Markov decision process (MDP) with the following main components. 
\begin{itemize}
  \item \emph{State}: The system state at $t$-th iteration observed by DO $n$ is the curator's previous payment, i.e., $s_n^t = p_n^{t-1}$.
  \item \emph{Action}: At $t$-th iteration, DO $n$ chooses a noise-saving action $a_n^{t} = \Delta\sigma_n^{t}$ with probability $\pi(s_n^t,a_n^t)$, where $\pi(.)$ is the mixed policy. Initially, $\pi(s_n^0,a_n^0) = 1/J$, $\forall n$.
  \item \emph{Reward}: The payoff $\mathcal{F}_{n}^t = \mathcal{F}_{n}({s}_n^t,{a}_n^t)$ defined in (\ref{eq:payoffDO}) serves as the immediate reward of DO $n$. The Q-function $\mathcal{Q}({s}_n^t,{a}_n^t)$ captures the expected long-term cumulative discounted reward of DO $n$ and is updated via the iteration Bellman equation, i.e.,
      \begin{align}\label{eq:Qfunction1}
        \mathcal{Q}({s}_n^t,{a}_n^t) \leftarrow & (1-\eta_1)\mathcal{Q}({s}_n^t,{a}_n^t) + \eta_1 \left[ \mathcal{F}_{n}^t \right. \nonumber \\
        & {+ \phi_1 \mathop {\max }\limits_{{a}_n^{t+1}} \mathcal{{Q}}\left({s}_n^{t+1}, {a}_n^{t+1}\right)} \Big],
      \end{align}
      where $\eta_1,\phi_1 \in (0,1]$ are the learning rate and discount factor, respectively. Initially, $\mathcal{Q}({s}_n^0,{a}_n^0)=0$, $\forall n$.
\end{itemize}

For better exploitation-exploration tradeoff, the mixed policy $\pi(s_n^t,a_n^t)$ in WoLF-PHC is updated by increasing the chance that acts greedily (i.e., attain the highest Q-value) by a small value $\psi_1$, and reducing other chances by $-{\psi_1}/{J}$:
\begin{align}\label{eq:mixPolicy1}
    \pi (s_n^t,a_n^t) \leftarrow \pi(s_n^t,a_n^t) \!+\! \left\{ \begin{array}{ll}
    \psi _1, ~{a}_n^t=\arg\max_{{a}_n}\mathcal{Q}({s}_n^t,{a}_n);\\
    -\frac{\psi_1}{J}, \,{\rm{otherwise}}.
    \end{array} \right.
\end{align}
Based on WoLF principle, $\psi_1$ is variable and has two values (i.e., $\psi_1^l$ and $\psi_1^w$ with $\psi_1^l>\psi_1^w$). The mixed policy is updated depending on whether the DO currently loses or wins, i.e.,
\begin{align}\label{eq:variablePolicy1}
    \psi_1 \!=\! \left\{ \begin{array}{ll}
    \psi_1^l, \,\sum\limits_{a_n} \pi(s_n^t,a_n)\mathcal{Q}({s}_n^t,{a}_n) \!\leq\! \sum\limits_{a_n} \overline{\pi}(s_n^t,a_n)\mathcal{Q}({s}_n^t,{a}_n);\\
    \psi_1^w, \,{\rm{otherwise}}.
    \end{array} \right.
\end{align}
The average mixed policy $\overline{\pi}(s_n,a_n)$ in Eq. (\ref{eq:variablePolicy1}) is updated by
\begin{align}\label{eq:avemixPolicy1}
    \overline{\pi}(s_n,a_n) \leftarrow \overline{\pi}(s_n,a_n) + \frac{{\pi}(s_n,a_n) - \overline{\pi}(s_n,a_n)}{\mathrm{count}(s_n)},
\end{align}
where $\mathrm{count}(s_n)$ denotes the times that state $s_n$ has been observed by DO $n$ until the current $t$-th interaction. 

\subsection{Intelligent Pricing Strategy Based on WoLF-PHC}\label{subsec:scheme2}
For the curator, its pricing strategy-making process under repeated interactions is formulated as a finite MDP as below. 
\begin{itemize}
  \item \emph{State}: The current system state at $t$-th iteration observed by the curator consists of the previous action vector of involved DOs, i.e., $\hat{\mathbf{s}}^t = [\hat{s}_n^t]_{n\in \mathcal{N}} =  [\Delta\sigma_n^{t-1}]_{n\in \mathcal{N}}$.
  \item \emph{Action}: At $t$-th iteration, the curator chooses a payment action $\hat{\mathbf{a}}^t = [\hat{a}_n^t]_{n\in \mathcal{N}} = [p_n^t]_{n\in \mathcal{N}}$ based on the mixed policy $\pi(\hat{\mathbf{s}}^t,\hat{\mathbf{a}}^t)$. Initially, $\pi(\hat{s}_n^0,\hat{a}_n^0) = 1/K$, $\forall n$.
  \item \emph{Reward}: The payoff $\mathcal{F}_{S}^t = \mathcal{F}_{S}(\hat{\mathbf{s}}^t,\hat{\mathbf{a}}^t)$ defined in (\ref{eq:payoffCurator}) is the curator's immediate reward. 
      The Q-function $\mathcal{Q}(\hat{\mathbf{s}}^t,\hat{\mathbf{a}}^t)$ denotes the curator's expected long-term cumulative discounted reward, which is updated by:
      \begin{align}\label{eq:Qfunction1}
\mathcal{Q}(\hat{s}_n^t,\hat{a}_n^t) \leftarrow & (1-\eta_2)\mathcal{Q}(\hat{s}_n^t,\hat{a}_n^t) + \eta_2 \left[ \mathcal{F}_{S,n}^t \right. \nonumber \\
& {+ \phi_2 \mathop {\max }\limits_{\hat{a}_n^{t+1}} \mathcal{{Q}}\left(\hat{s}_n^{t+1}, \hat{a}_n^{t+1}\right)} \Big],
\end{align}
where $\eta_2,\phi_2 \in (0,1]$ are the learning rate and discount factor, respectively. Initially, $\mathcal{Q}(\hat{s}_n^0,\hat{a}_n^0)=0$, $\forall n$.
\end{itemize}

Similarly, the mixed policy $\pi(\hat{\mathbf{s}}^t,\hat{\mathbf{a}}^t)$ is updated by:
\begin{align}\label{eq:mixPolicy2}
    \pi (\hat{s}_n^t,\hat{a}_n^t) \leftarrow \pi(\hat{s}_n^t,\hat{a}_n^t) \!+\! \left\{ \begin{array}{ll}
    \psi _2, ~\hat{a}_n^t=\arg\max_{\hat{a}_n}\mathcal{Q}(\hat{s}_n^t,\hat{a}_n);\\
    -\frac{\psi_2}{K}, \,{\rm{otherwise}}.
    \end{array} \right.
\end{align}
In (\ref{eq:mixPolicy2}), the variable $\psi_2$ has two values (i.e., $\psi_2^l$ and $\psi_2^w$ with $\psi_2^l>\psi_2^w$) based on WoLF principle, which is determined by
\begin{align}\label{eq:variablePolicy2}
    \psi_2 \!=\! \left\{ \begin{array}{ll}
    \psi_2^l, \,\sum\limits_{\hat{a}_n} \pi(\hat{a}_n^t,\hat{a}_n)\mathcal{Q}(\hat{s}_n^t,\hat{a}_n) \!\leq\! \sum\limits_{\hat{a}_n} \overline{\pi}(\hat{s}_n^t,\hat{a}_n)\mathcal{Q}(\hat{s}_n^t,\hat{a}_n);\\
    \psi_2^w, \,{\rm{otherwise}}.
    \end{array} \right.
\end{align}
The average mixed policy $\overline{\pi}(\hat{s}_n,\hat{a}_n)$ in Eq. (\ref{eq:variablePolicy2}) is updated similar to that in (\ref{eq:avemixPolicy1}).

\emph{Remark.} The time complexity of the proposed two-layer RL-based approach yields $\mathcal{O}(N K J)$, and its convergence is validated using experiments in the next section.

\section{PERFORMANCE EVALUATION}\label{sec:SIMULATION}

\subsection{Experiment Setup}\label{subsec:evalution1}
\textbf{Datasets and Models.} The classic MNIST dataset is used to perform handwritten digits recognition tasks among $N=100$ DOs under FL. 
For dataset partition among DOs, the Dirichlet parameter $\beta$, which controls the non-IID degree of DO's training samples, is chosen within $[0.05,20]$. Each DO applies the 4-layer CNN model to compute the local model with batch size $64$, learning rate $0.05$, and local epoch $1$. The maximum communication round is set as $I=30$.

\textbf{DyPP Game.} For Gaussian noise adding, we set $\sigma_{\max}=0.6$. For quality-loss mapping, we set $\zeta_1=35.4278$, $\zeta_2=102.2444$. For the payoff model, we set $\varpi=0.6$, $\lambda_s = 0.2$, $\lambda_r = 0.08$, $\mu = 0.13$, $v = 2.5$, $p_{\max} = 16$, $c_n \in [0.5,4]$. For the WoLF-PHC model, we set $\phi_1 = \phi_2 =0.8$, $\psi_1^w = \psi_2^w = 1 / (50 + t/50)$, $\psi_1^l  = \psi_2^l  = 2 \psi_1^w$, $K=32$, $J=12$. According to \cite{corr/abs-2212-13992}, the model loss function can be well-fitted by the 3D sigmoid curve with curve-fitting parameters $\mu_1=0.013$, $\mu_2=0.0044$, $\mu_3=0.0057$, $\mu_4=8.18$, $\mu_5=0.14$.
We compare the proposed approach with the following benchmarks.
\begin{itemize}
  \item In \textbf{two-layer Q-learning scheme}, both the DO and curator employ Q-learning to obtain their optimal policies.
  \item In \textbf{greedy scheme}, both the DO and curator behave greedily in the repeated DyPP game.
\end{itemize}


\begin{figure}[!t]\vspace{-0.2cm}
\setlength{\abovecaptionskip}{-0.1cm}
\begin{minipage}[t]{0.25\textwidth}
\centering
    \includegraphics[height=3.5cm,width=\textwidth]{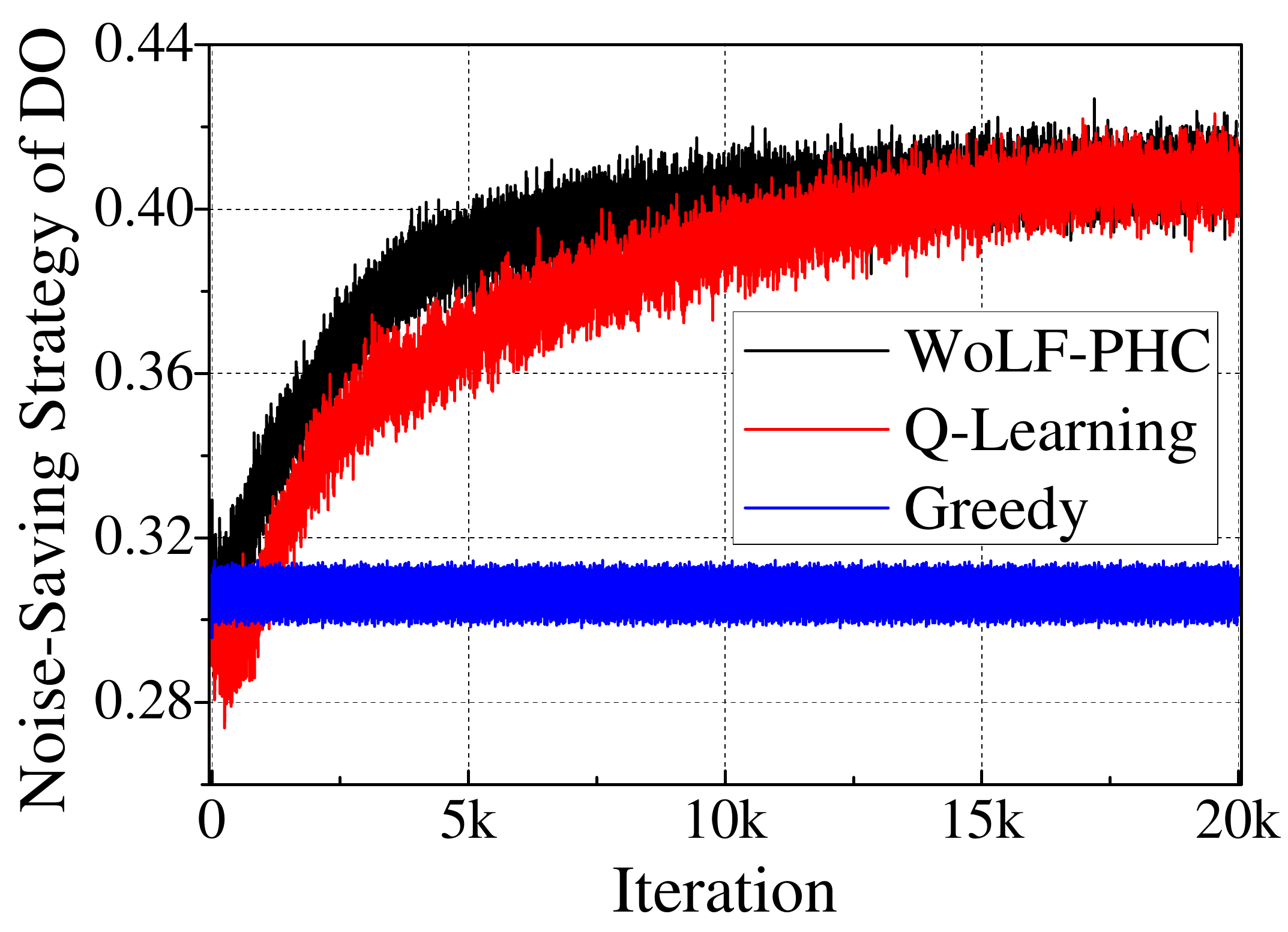}
    \caption{Convergence of DO's noise-saving strategy using WoLF-PHC, compared with Q-learning and greedy schemes.}\label{fig:simu2}
\end{minipage}~
\begin{minipage}[t]{0.25\textwidth}
\centering
    \includegraphics[height=3.5cm,width=\textwidth]{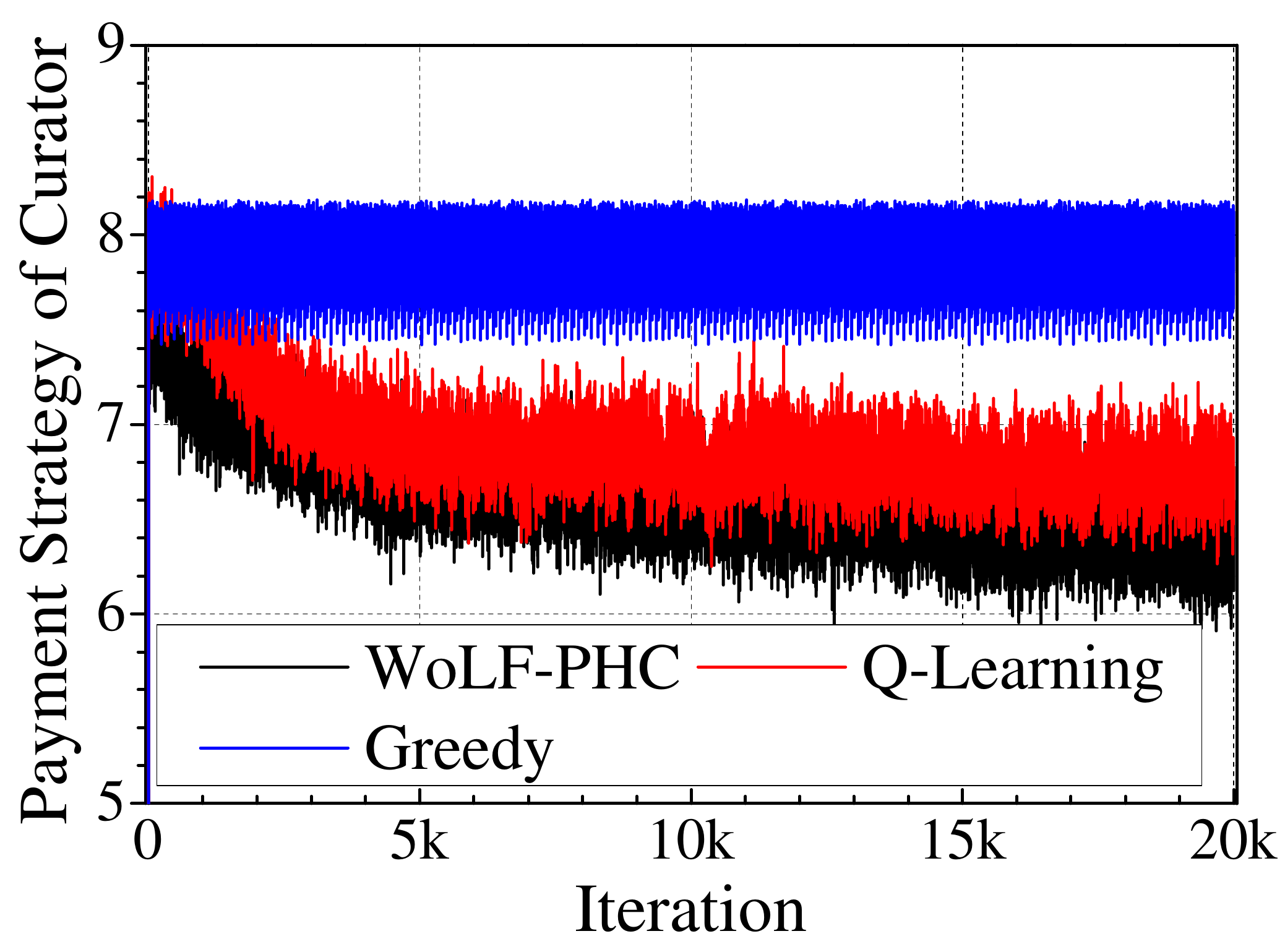}
    \caption{Convergence of curator's payment strategy using WoLF-PHC, compared with Q-learning and greedy schemes.}\label{fig:simu3}
\end{minipage}
\end{figure}

\subsection{Experimental Results}\label{subsec:evalution2}

In Figs.~\ref{fig:simu2} and \ref{fig:simu3}, we validate the convergence of our two-layer RL approach in solving the DyPP game.
These two figures show that our two-layer WoLF-PHC scheme can accelerate the convergence rate and improve model utility, compared with the two-layer Q-learning and greedy schemes. 
Among the three approaches, the greedy scheme has the fastest convergence rate but suffers the worst model performance. The two-layer Q-learning requires the largest iterations to obtain the optimal policy for both sides, and its slow convergence may lower DOs' willingness to join DyPP game to trade privacy for utility in FL.
Besides, in Fig.~\ref{fig:simu2}, the saved DP noise scale keeps increasing before attaining a stable value, while the corresponding payment in Fig.~\ref{fig:simu3} keeps decreasing before it converges to the stable state.
The reason is that the initial high compensation motivates DOs to gradually increase their saved noise scales by adding Gaussian noise with smaller variance for higher payoffs. Meanwhile, after observing DOs' high noise-saving actions, the curator intends to gradually reduce its payment for enhanced payoff.

\section{Conclusion}\label{sec:CONSLUSION}
For better privacy-utility tradeoff in practical DP-based FL services, this paper has proposed a novel DyPP game approach that allows DOs to trade individual privacy (i.e., determining the local noise-saving strategy) for improved global model utility by providing differentiated payment contracts to compensate DOs' privacy losses.
In the DyPP game, the multi-dimensional information asymmetry between DOs and the curator, as well as their varying private information under distinct FL tasks, poses a challenge to derive the closed-form expression of the NE. We have also devised a fast RL algorithm to enable both DOs and the curator to quickly learn their optimal policies and adapt to dynamic and uncertain environments without being aware of the player's private information.
Experimental results have shown that the proposed scheme attains a faster convergence rate and enhanced model utility with lower payments, compared with benchmarks.
For future work, the DyPP game with bounded rationality and continuous action space in FL will be investigated.

\section*{Acknowledgment}
This work was supported in part by NSFC (nos. U22A2029, U20A20175), and the Fundamental Research Funds for the Central Universities.

\bibliographystyle{IEEETran}
\bibliography{ref}
\end{document}